\documentstyle[11pt,aasms4]{article}

\def\vsini{$v\sin i$}

\begin{document}

\title{The Visual Orbit and Evolutionary State of 12~Bo\"{o}tes}
\author{A.F.~Boden\altaffilmark{1,2},
	M.J.~Creech-Eakman\altaffilmark{3,2},
	D.~Queloz\altaffilmark{4,2}
}
\altaffiltext{1}{Infrared Processing and Analysis Center, California Institute of Technology}
\altaffiltext{2}{Jet Propulsion Laboratory, California Institute of Technology}
\altaffiltext{3}{Geology and Planetary Sciences, California Institute of Technology}
\altaffiltext{4}{Observatoire de Gen\`{e}ve}
\authoremail{bode@huey.jpl.nasa.gov}

\begin{abstract}
We report on the determination of the visual orbit of the double-lined
spectroscopic binary system 12~Bo\"{o}tes with data obtained by the
Palomar Testbed Interferometer in 1998 and 1999.  12~Boo is a nearly
equal-mass double-lined binary system whose spectroscopic orbit is
well known.  We have estimated the visual orbit of 12~Boo from our
interferometric visibility data fit both separately and in conjunction
with archival and CORAVEL radial velocity data.  Our 12~Boo orbit is
in good agreement with the spectroscopic results, and the physical
parameters implied by a combined fit to our visibility data and radial
velocity data result in precise component masses.  In particular, the
orbital parallax of the system is determined to be 27.09 $\pm$ 0.41
mas, and masses of the two components are determined to be 1.435 $\pm$
0.023 M$_{\sun}$ and 1.409 $\pm$ 0.020 M$_{\sun}$, respectively.

Somewhat remarkably, even though the two components are nearly equal
mass, the system exhibits a significant brightness difference between
the components in the near infrared and visible.  We attribute this
brightness difference to evolutionary differences between the two
components in their transition between main sequence and giant
evolutionary phases, and based on theoretical isochrones we can
estimate a system age.  Further, because the atmospheres of the two
components are becoming more convective, we suggest the system
components are currently at or near synchronous rotation, and the
system orbit is in the process of circularizing.

\end{abstract}

\section{Introduction}

12~Bo\"{o}tes (d~Bo\"{o}tes, HR 5304, HD 123999) is a short-period
(9.6 d) binary system with nearly-equal mass ($q \sim$ 0.98)
components.  The system was first detected as a radial velocity
variable by Campbell \& Wright (1900), and the first ``good''
double-lined orbit was calculated by Abt \& Levy (1976, hereafter
AL76).  Merrill (1922) attempted to resolve the system with the Mount
Wilson interferometer, but did not see any visibility variations and
placed it in his ``Apparently Single'' category.  Very recently the
AL76 orbit has been reconfirmed by an independent CORAVEL radial
velocity orbit (\cite{DeMedeiros99}, hereafter DU99).  The composite
system has been consistently assigned the spectral type F8IV -- F9IVw,
the latter by Barry (1970), with the ``w'' indicating weak ultraviolet
metallic features.  There is general consensus that the components of
12~Boo have evolved off the main sequence.  All studies seem to
confirm that 12~Boo has heavy element abundances near solar
proportions (\cite{Duncan81,Balachandran90,Lebre99}).

12~Boo is listed as a triple system by Tokovinin (1997), presumably
because the WDS lists 12~Boo as having a visual companion at a
separation of approximately 1'' at a position angle of 8$^{\circ}$,
but lists no magnitude (\cite{WDS97}).  However, McAlister, Hartkopf,
\& Mason (1992) find no companion to 12~Boo within the limits of
their speckle observations (separation greater than 0.03'' and
$\Delta$ m $<$ 1.5 mag), and therefore list it as ``single'' in their
Table 5.

Herein we report the determination of the 12~Boo visual orbit from
near-infrared, long-baseline interferometric measurements taken with
the Palomar Testbed Interferometer (PTI).  PTI is a 110-m $H$
(1.6$\mu$m) and $K$-band (2.2$\mu$m) interferometer located at
Palomar Observatory, and described in detail elsewhere
(\cite{Colavita99a}).  PTI has a minimum fringe spacing of roughly 4
milliarcseconds (10$^{-3}$ arcseconds, mas) in $K$-band at the sky
position of 12~Boo, allowing resolution of this binary system.  We
further add photometric and spectroscopic measurements in an attempt
to understand the fundamental stellar parameters and evolution of the
12~Boo components.

\section{Observations}
\label{sec:observations}
The interferometric observable used for these measurements is the
fringe contrast or {\em visibility} (squared) of an observed
brightness distribution on the sky.  Normalized in the interval [0:1],
a single star exhibits monochromatic visibility modulus in a uniform
disk model given by:
\begin{equation}
V =
\frac{2 \; J_{1}(\pi B \theta / \lambda)}{\pi B \theta / \lambda}
\label{eq:V_single}
\end{equation}
where $J_{1}$ is the first-order Bessel function, $B$ is the projected
baseline vector magnitude at the star position, $\theta$ is the
apparent angular diameter of the star, and $\lambda$ is the wavelength
of the interferometric observation.  The expected squared visibility
in a narrow bandpass for a binary star such as 12~Boo is given by:
\begin{equation}
V_{nb}^{2} = \frac{V_{1}^2 + V_{2}^2 \; r^2
	        + 2 \; V_{1} \; V_{2} \; r \;
	          \cos(\frac{2 \pi}{\lambda} \; {\bf {B}} \cdot {\bf {s}})}
	      {(1 + r)^2}
\label{eq:V2_double}
\end{equation}
where $V_{1}$ and $V_{2}$ are the visibility moduli for the two stars
separately as given by Eq.~\ref{eq:V_single}, $r$ is the apparent
brightness ratio between the primary and companion, ${\bf {B}}$ is the
projected baseline vector at the system sky position, and ${\bf {s}}$
is the primary-secondary angular separation vector on the plane of the
sky (\cite{Hummel95}).  The $V^2$ observables used in our 12~Boo study
are both narrow-band $V^2$ from individual spectral channels
(\cite{Colavita99a}), and a synthetic wide-band $V^2$, given by an
incoherent SNR-weighted average $V^2$ of the narrow-band channels in
the PTI spectrometer (\cite{Colavita99b}).  In this model the expected
wide-band $V^2_{wb}$ observable is approximately given by an average
of the narrow-band formula over the finite bandpass of the
spectrometer:
\begin{equation}
V^{2}_{wb} = \frac{1}{n}\sum_{i}^{n} V^{2}_{nb-i}(\lambda_i)
\label{eq:V2_doubleWB}
\end{equation}
where the sum runs over the channels covering the infrared $H$-band
(1.5 -- 1.8 $\mu$m) and $K$-band (2 - 2.4 $\mu$m) of the PTI
spectrometer; PTI operating wavebands are excellent matches to the CIT
photometric system (\cite{Colavita99a,Elias82,Elias83}).  Separate
calibrations and binary model fits to the narrow-band and synthetic
wide-band $V^2$ datasets yield statistically consistent results, with
the synthetic wide-band data exhibiting superior fit performance.
Consequently we will present only the results from the synthetic
wide-band data.

12~Boo was observed in conjunction with objects in our calibrator list
by PTI in $K$-band ($\lambda \sim 2.2 \mu$m) on 17 nights between 21
June 1998 and 28 June 1999, covering roughly 39 periods of the system.
Additionally, 12~Boo was observed by PTI in $H$-band ($\lambda \sim
1.6 \mu$m) on 28 May 1999.  12~Boo, along with calibration objects,
was observed multiple times during each of these nights, and each
observation, or scan, was approximately 130 sec long.  For each scan
we computed a mean $V^2$ value from the scan data, and the error in
the $V^2$ estimate from the rms internal scatter (\cite{Colavita99b}).
12~Boo was always observed in combination with one or more calibration
sources within $\sim$ 10$^{\circ}$ on the sky.  For our study we have
used three stars as calibration objects: HD 121107 (G5 III), HD 128167
(F2 V), and HD 123612 (K5 III).  Table \ref{tab:calibrators} lists the
relevant physical parameters for the calibration objects.

\begin{table}[t]
\begin{center}
\begin{small}
\begin{tabular}{|c|c|c|c|c|}
\hline
Object    & Spectral & Star        & 12~Boo         & Adopted Model \\
Name      & Type     & Magnitude   & Separation     & Diameter (mas)  \\
\hline
HD 121107 & G5 III   & 6.1 V/4.1 K & 8.2$^{\circ}$  & 0.84 $\pm$ 0.06   \\
HD 128167 & F2 V     & 5.9 V/3.9 K & 7.1$^{\circ}$  & 0.77 $\pm$ 0.04   \\
HD 123612 & K5 III   & 5.7 V/3.5 K & 0.92$^{\circ}$ & 1.33 $\pm$ 0.10   \\
\hline
\end{tabular}
\caption{PTI 12~Boo Calibration Objects Considered in our Analysis.
The relevant parameters for our three calibration objects are
summarized.  The apparent diameter values are determined from
effective temperature and bolometric flux estimates based on archival
broad-band photometry, and visibility measurements with PTI.
\label{tab:calibrators}}
\end{small}
\end{center}
\end{table}

The calibration of 12~Boo $V^2$ data is performed by estimating the
interferometer system visibility ($V^{2}_{sys}$) using calibration
sources with model angular diameters, and then normalizing the raw
12~Boo visibility by $V^{2}_{sys}$ to estimate the $V^2$ measured by
an ideal interferometer at that epoch (\cite{Mozurkewich91,Boden98}).
Calibrating our 12~Boo dataset with respect to the three calibration
objects listed in Table \ref{tab:calibrators} results in a total of 72
calibrated scans (62 in $K$, 10 in $H$) on 12~Boo over 18 nights in
1998 and 1999.  Our calibrated synthetic wide-band $V^2$ measurements
are summarized in Table~\ref{tab:dataTable}.

\begin{table}
\dummytable\label{tab:dataTable}
\end{table}

\begin{table}
\dummytable\label{tab:RVdata}
\end{table}

To our PTI visibilities and 17 double-lined AL76 radial velocity
measurements we have added seven double-lined radial velocity
measurements from CORAVEL.  These seven CORAVEL RV measurements are a
subset of the 12 measurements from DU99.  The 24 radial velocity
measurements used in our analysis are summarized in Table
\ref{tab:RVdata}.

Finally, on 29 and 30 March 1999 we obtained broad-band infrared
photometry on the 12~Boo system using the 200'' Hale telescope at
Palomar Observatory.  12~Boo was observed during photometric
conditions with respect to Elias IR standards 9529 and 9539 (HD 105601
and HD 129653 respectively, \cite{Elias82}) at airmasses less than
1.2.  Table~\ref{tab:photometry} summarizes the results of our
photometric measurements on 12~Boo.

\begin{table}[t]
\begin{center}
\begin{small}
\begin{tabular}{|c|c|}
\hline
Band       & Magnitude   \\
\hline
$J_{\rm CIT}$  & 3.810 $\pm$ 0.020  \\
$H_{\rm CIT}$  & 3.600 $\pm$ 0.020  \\
$K_{\rm CIT}$  & 3.550 $\pm$ 0.016  \\
\hline
$V$ (archival) & 4.83 $\pm$ 0.01 \\
\hline
\end{tabular}
\caption{12~Boo Near-Infrared Photometric Measurements.  We summarize
our near-infrared photometric measurements on the 12~Boo system (taken
at the 200'' telescope at Palomar Observatory on 29 and 30 March
1999), and archival $V$-band photometry from the Simbad database.  Our
infrared photometry is taken in the CIT system
(\cite{Elias82,Elias83}).
\label{tab:photometry}}
\end{small}
\end{center}
\end{table}

\section{Orbit Determination}

As in previous papers in this series (\cite{Boden99a,Boden99b}) the
estimation of the 12~Boo visual orbit is made by fitting a Keplerian
orbit model with visibilities predicted by Eqs.~\ref{eq:V2_double} and
\ref{eq:V2_doubleWB} directly to the calibrated (narrow-band and
synthetic wide-band) $V^2$ data on 12~Boo (see also
\cite{Armstrong92b,Hummel93}, 1995, 1998).  The fit is non-linear in
the Keplerian orbital elements, and is therefore performed by
non-linear least-squares methods (i.e.~the Marquardt-Levenberg method,
\cite{Press92}).  As such, this fitting procedure takes an initial
estimate of the orbital elements and other parameters (e.g. component
angular diameters, brightness ratio), and evolves that model into a
new parameter set that best fits the data.  However, the chi-squared
surface has many local minima in addition to the global minimum
corresponding to the true orbit.  Because the Marquardt-Levenberg
method strictly follows a downhill path in the $\chi^2$ manifold, it
is necessary to thoroughly survey the space of possible binary
parameters to distinguish between local minima and the true global
minimum.  In addition, as the $V^2$ observable for the binary
(Eqs.~\ref{eq:V2_double} and \ref{eq:V2_doubleWB}) is invariant under
a rotation of 180$^{\circ}$, we cannot differentiate between an
apparent primary/secondary relative orientation and its mirror image
on the sky.  Consequently there remains a 180$^{\circ}$ ambiguity in
our determination of the longitude of the ascending node, $\Omega$,
which we quote by convention in the interval [0:180).  By similar
arguments our $V^2$ observable does not distinguish the longitude of
periastron ($\omega$) for the primary and secondary component.  We
have constrained our estimate to be grossly (within 180$^{\circ}$)
consistent with the $\omega_1$ value of approximately 290$^{\circ}$
(AL76, DU99).

In addition to our PTI visibility data we have used the double-lined
radial velocity data from AL76 and CORAVEL.  We incorporate these data
into the orbit estimation, utilizing both interferometric visibility
and radial velocity data either separately or simultaneously
(\cite{Hummel98,Boden99b}).  The $\omega$-degeneracy discussed above
is resolved by the inclusion of radial velocity data in our orbital
solution (Table~\ref{tab:orbit}), however the determination of
$\Omega$ remains ambiguous by 180$^\circ$.

In the case of 12~Boo the parameter space is significantly narrowed by
the high-quality spectroscopic orbits from AL76 and DU99, and the
Hipparcos distance determination sets the rough scale of the
semi-major axis (\cite{HIP97,Perryman97}).  Further, at the distance
of 12~Boo the apparent diameters of the two components of the system
are not strongly resolved by PTI, so we have constrained the estimated
diameters of both components to model values of 0.63 $\pm$ 0.06 and
0.46 $\pm$ 0.05 mas for the primary and secondary components
respectively (see the discussion in \S~\ref{sec:physics}).  Given this
limited parameter space, the correct orbit solution is readily
obtained by exhaustive search for the global minimum in the $\chi^2$
manifold.

\begin{figure}
\epsscale{0.7}
\plotone{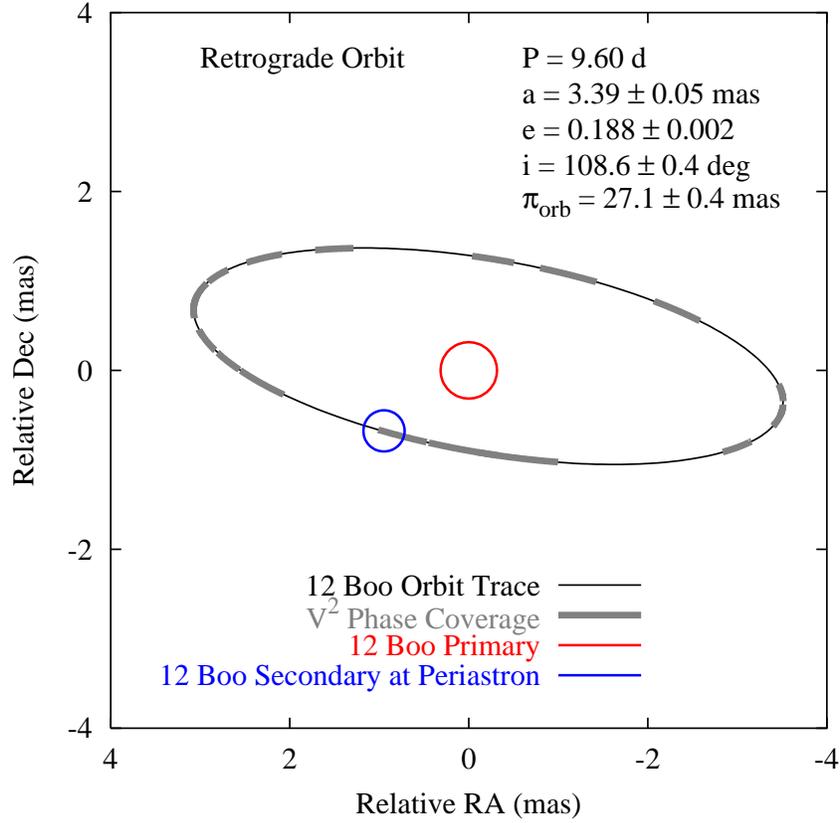}
\caption{Visual Orbit of 12~Boo.  The relative visual orbit model of
12~Boo is shown, with the primary and secondary objects rendered at
T$_0$ (periastron).  The heavy lines along the relative orbit indicate
areas where we have orbital phase coverage in our PTI data (they are
not separation vector estimates); our data sample most phases of the
orbit well, leading to a reliable orbit determination.  Component
diameter values are estimated (see discussion in
\S~\ref{sec:physics}), and are rendered to scale.
\label{fig:12boo_orbit}}
\end{figure}

Figure \ref{fig:12boo_orbit} depicts the relative visual orbit of the
12~Boo system, with the primary component rendered at the origin, and
the secondary component rendered at periastron.  We have indicated the
phase coverage of our $V^2$ data on the relative orbit with heavy
lines; our data samples most phases of the orbit well, leading to a
reliable orbit determination.

\begin{figure}
\epsscale{0.8}
\plotone{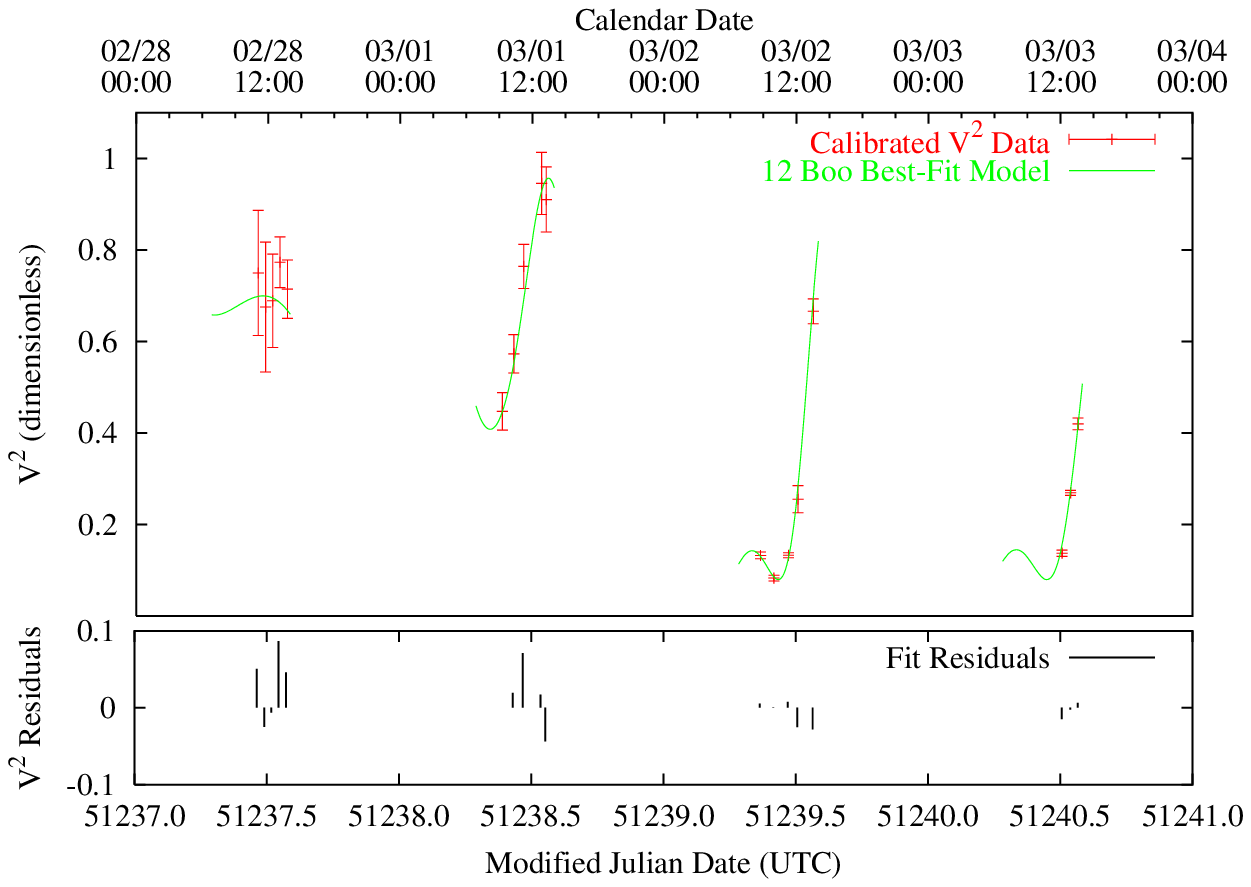}\\
\plotone{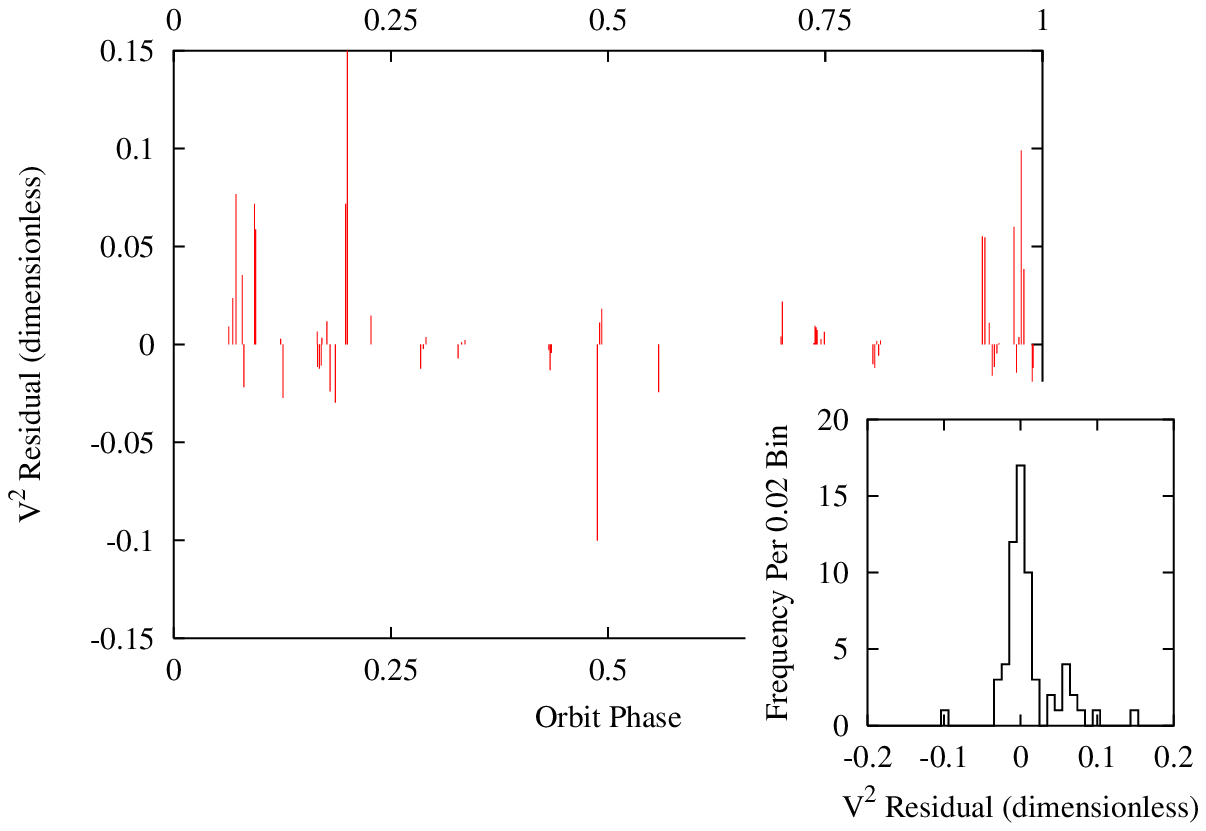}
\caption{$V^2$ Fit of 12~Boo.  a: Four consecutive nights (28 February
-- 3 March 1999) of calibrated $V^2$ data on 12~Boo, and $V^2$
predictions from the best-fit model for the system.  The lower frame
shows individual $V^2$ residuals between the calibrated data and
best-fit model.  b: A phase plot of $K$-band $V^2$ fit residuals from
our Full-Fit solution (Table~\ref{tab:orbit}).  We have inset a $V^2$
error residual histogram.
\label{fig:12boo_V2fit}}
\end{figure}

\begin{figure}
\epsscale{1.0}
\plotone{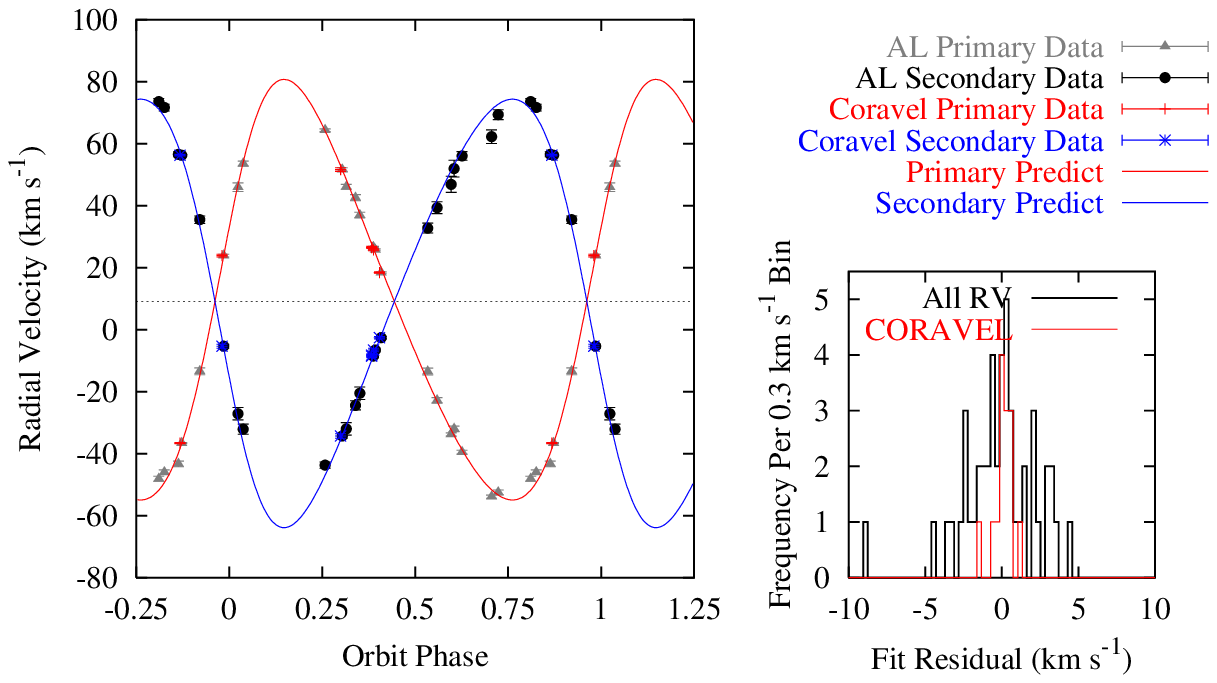}
\caption{RV Fit of 12~Boo.  A phase plot of radial velocity data from
AL76 and CORAVEL and fit predictions from our Full-Fit solution
(Table~\ref{tab:orbit}).  We have inset RV error residual histograms
for all the RV data, and separately for the DU99 CORAVEL RV; the
higher precision of the CORAVEL data is evident.
\label{fig:12boo_RVfit}}
\end{figure}

Table \ref{tab:dataTable} lists the complete set of $V^2$ measurements
in our 12~Boo dataset and the prediction based on the best-fit orbit
model (our ``Full-Fit'' model, Table \ref{tab:orbit}) for 12~Boo.
Table~\ref{tab:dataTable} gives $V^2$ measurements and times,
measurement errors, model predictions, the photon-weighted average
wavelength, $u-v$ coordinates, and on-target hour angle for each of
our calibrated 12~Boo observations.  Figures \ref{fig:12boo_V2fit} and
\ref{fig:12boo_RVfit} illustrate our model fit for 12~Boo.  Figure
\ref{fig:12boo_V2fit}a shows four consecutive nights of PTI $V^2$ data
on 12~Boo (28 Feb -- 3 Mar 1999), and $V^2$ predictions based on the
best-fit model for the system (our ``Full-Fit'' model, Table
\ref{tab:orbit}).  Figure \ref{fig:12boo_V2fit}b gives a phase plot of
$V^2$ residuals, with an inset $V^2$ residual histogram.  The model
predictions are in good agreement with the observed data, with an rms
$V^2$ residual of 0.033 (average absolute $V^2$ residual of 0.023),
and a $\chi^2$ per Degree of Freedom (DOF) of 1.04.  The quality of
the $V^2$ fit is similar to those seen in other PTI orbital analyses,
and we can see no signs of either bias or excess noise contributed by
a putative optical companion at a separation of 1''.  Figure
\ref{fig:12boo_RVfit} gives a radial velocity phase plot of the AL76
and CORAVEL radial velocity data and the predictions of our
``Full-Fit'' orbital solution.  The RV rms residual in our solution is
2.3 km s$^{-1}$ (average absolute residual of 1.7 km s$^{-1}$), but
the fit quality of the CORAVEL RV subset (Table~\ref{tab:RVdata})
is considerably better, with an rms residual of 0.59 km s$^{-1}$
(average absolute residual of 0.44 km s$^{-1}$).

Spectroscopic orbit parameters (from AL76 and DU99) and our visual and
spectroscopic orbit parameters of the 12~Boo system are summarized in
Table \ref{tab:orbit}.  We give the results of separate fits to only
our $V^2$ data (our ``$V^2$-only Fit'' solution), and a simultaneous
fit to our $V^2$ data and the double-lined radial velocities from AL76
and CORAVEL (our ``Full-Fit'' solution) -- both with component
diameters constrained as noted above.  For the orbit parameters we
have estimated from our visibility data we list a total one-sigma
error in the parameter estimate, and the separate one-sigma errors in
the parameter estimates from statistical (measurement uncertainty) and
systematic error sources.  In our analysis the dominant forms of
systematic error are: (1) uncertainties in the calibrator angular
diameters (Table \ref{tab:calibrators}); (2) uncertainty in the
center-band operating wavelength ($\lambda_0 \approx$ 2.2 $\mu$m),
taken to be 20 nm ($\sim$1\%); (3) the geometrical uncertainty in our
interferometric baseline ( $<$ 0.01\%); and (4) uncertainties in
orbital parameters constrained in our fitting procedure (i.e.~the
angular diameters in both solutions, the period in the ``$V^2$-only''
solution).

\begin{table}
\begin{center}
\begin{small}
\begin{tabular}{ccc|cc}
\hline
Orbital			& AL76                & DU99   & \multicolumn{2}{c}{PTI 98/99} \\
\cline{4-5}
Parameter       	&                     &        & $V^2$-only Fit   & Full Fit \\
\hline \hline
Period (d)              & 9.604538            & 9.6046 & {\em 9.604565}   & 9.604565       \\
                        & $\pm$ 2.2 $\times$ 10$^{-5}$ & $\pm$ 1 $\times$ 10$^{-4}$ &   & $\pm$ 1.0 (1.0/0.07) $\times$ 10$^{-5}$ \\
T$_{0}$ (MJD)           & 17679.511           & 48990.29 & 51237.749        & 51237.779  \\
                        & $\pm$ 0.084         & $\pm$ 0.03 & $\pm$ 0.024 (0.016/0.018) &  $\pm$ 0.020 (0.018/0.008) \\
$e$                     & 0.1933              & 0.193  &   0.1781         & 0.1884 \\
			& $\pm$ 0.0080        & $\pm$ 0.004 & $\pm$ 4.8 (4.4/2.0) $\times$ 10$^{-3}$ & $\pm$ 2.2 (2.2/0.2) $\times$ 10$^{-3}$ \\
K$_1$ (km s$^{-1}$)     & 67.4 $\pm$ 0.8      & 67.11 $\pm$ 0.41 &                  & 67.84 $\pm$ 0.31 (0.31/0.03) \\
K$_2$ (km s$^{-1}$)     & 66.5 $\pm$ 0.9      & 70.02 $\pm$ 0.48 &                  & 69.12 $\pm$ 0.48 (0.48/0.03) \\
$\gamma$ (km s$^{-1}$)  & 9.2 $\pm$ 0.4       & 9.29 $\pm$ 0.19  &                  & 9.11 $\pm$ 0.13 (0.13/0.01) \\
$\omega_{1}$ (deg)      & 290 $\pm$ 3         & 286.19 $\pm$ 1.31 &  287.0 $\pm$ 1.3 (0.9/0.9) & 287.03 $\pm$ 0.75 (0.69/0.30) \\
$\Omega_{1}$ (deg)      &                     &                   & 9.56 $\pm$ 0.41 (0.40/0.07) & 10.17 $\pm$ 0.45 (0.40/0.21) \\
$i$ (deg)               &                     &                   & 108.84 $\pm$ 0.32 (0.19/0.26) & 108.58 $\pm$ 0.36 (0.29/0.21) \\
$a$ (mas)               &                     &                   & 3.413 $\pm$ 0.039 (0.028/0.027) & 3.392 $\pm$ 0.050 (0.036/0.034) \\
$\Delta K_{\rm CIT}$ (mag)        &                     &                   & 0.614 $\pm$ 0.015 (0.011/0.011) & 0.618 $\pm$ 0.022 (0.019/0.011) \\
$\Delta H_{\rm CIT}$ (mag)        &                     &                   & 0.588 $\pm$ 0.066 (0.063/0.019) & 0.566 $\pm$ 0.066 (0.063/0.019) \\
$\chi^2$/DOF	        &                     &                   & 0.82             & 1.2 (1.04 $V^2$/2.8 RV)  \\
$\overline{|R_{V^2}|}$  &                     &                   & 0.023 & 0.023 \\
$\overline{|R_{RV}|}$ (km s$^{-1}$)  & 1.4    & 0.90              &                  & 1.7 (2.2 AL/0.44 COR)  \\
\hline
\end{tabular}
\end{small}
\caption{Orbital Parameters for 12~Boo.  Summarized here are the
apparent orbital parameters for the 12~Boo system as determined by
AL76, DU99, and PTI.  We give two separate fits to our data, with and
without including the double-lined AL76 and CORAVEL radial velocities
in the fit.  For parameters we have estimated by including our PTI
observations we separately quote one-sigma errors from both
statistical and systematic sources, given as
($\sigma_{stat}$/$\sigma_{sys}$), and the total error as the sum of
the two in quadrature.  Quantities given in italics are constrained to
the listed values in our model fits.  We have quoted the longitude of
the ascending node parameter ($\Omega$) as the angle between local
East and the orbital line of nodes measured positive in the direction
of local North.  Due to the degeneracy in our $V^2$ observable there
is a 180$^\circ$ ambiguity in $\Omega$; by convention we quote it in
the interval of [0:180).  We quote mean absolute $V^2$ and RV
residuals in the fits, $\overline{|R_{V^2}|}$ and
$\overline{|R_{RV}|}$ respectively.
\label{tab:orbit}}
\end{center}
\end{table}

\section{Physical Parameters}
\label{sec:physics}
Physical parameters derived from our 12~Boo ``Full-Fit''
visual/spectroscopic orbit are summarized in Table \ref{tab:physics}.
As in Table \ref{tab:orbit}, for physical parameters we have estimated
we quote total one-sigma errors, and statistical and systematic
contributions.  (Exceptions to this are quantities we have estimated
using the $V$-band spectroscopy discussed in \S~\ref{sec:tidal}; the
error is taken as statistical, and is relatively large compared to the
interferometric determinations.)  Notable among these is the
high-precision determination of the component masses for the system, a
virtue of the precision of the AL76 and CORAVEL radial velocity
measurements on both components and the moderately high inclination of
the orbit.  We estimate the masses of the primary and secondary
components as 1.435 $\pm$ 0.023 and 1.408 $\pm$ 0.020 M$_{\sun}$,
respectively.

The Hipparcos catalog lists the parallax of 12~Boo as 27.27 $\pm$ 0.78
mas (\cite{HIP97}).  The distance determination to 12~Boo based on our
orbital solution is 36.93 $\pm$ 0.56 pc, corresponding to an orbital
parallax of 27.08 $\pm$ 0.41 mas, consistent with the Hipparcos result
at 0.7\% and 0.2-sigma.

\paragraph{Component Diameters and Effective Temperatures}
The ``effective'' net angular diameter of the 12~Boo system has been
estimated using the infrared flux method (IRFM) by Blackwell and
collaborators (\cite{Blackwell90,Blackwell94}) at approximately 0.8
mas.  At this size neither of the 12~Boo components are resolved by
PTI, and we must resort to model diameters for the components.
Following Blackwell, we have estimated 12~Boo component diameters
through bolometric flux and effective temperature ($T_{\rm eff}$)
arguments.  Blackwell and Lynas-Gray (1994) list the bolometric flux
of the 12~Boo system at 3.11$\times$10$^{-10}$ W m$^{-2}$, and $T_{\rm
eff}$ as 6204 K, both without error estimates.  By assuming our
$K$-band flux ratio as a surrogate for the bolometric flux ratio
(correct in the limit of similar $T_{\rm eff}$ for the two components)
it is straightforward to compute component diameters as a function of
individual component $T_{\rm eff}$.  If we assume the 6204 K $T_{\rm
eff}$ for both components we arrive at diameter estimates of 0.62 and
0.46 mas for the primary and secondary components respectively.
However, comparing our interferometric component magnitude difference
in $K$-band and a spectroscopic estimate in $V$-band as we shall see
in \S~\ref{sec:tidal}, we can employ an empirical effective
temperature -- ($V$ - $K_n$) color index relationship for subgiant
stars derived by Blackwell et al.~(1990) to estimate $T_{\rm eff}$ of
6050 $\pm$ 75 and 6250 $\pm$ 130 K for the more evolved primary and
less-evolved secondary components respectively.  (In this computation
we have corrected for the increase in the 12~Boo $T_{\rm eff}$
estimate between \cite{Blackwell90} and \cite{Blackwell94}, and have
used the ($V$ - $K$) to ($V$ - $K_n$) empirical color index correction
of \cite{Selby88} with the component ($V$ - $K$) values of
Table~\ref{tab:physics}.)  With these individual component $T_{\rm
eff}$ we find model diameters of 0.65 and 0.46 mas for the primary and
secondary components respectively, similar to the values derived from
applying the Blackwell $T_{\rm eff}$ to both components.  We have
averaged these two sets of diameter estimates for our final model
diameters, arriving at 0.63 and 0.46 mas for the primary and secondary
component angular diameters respectively.  Rigorous error estimates in
these diameter models are made impossible without estimates of the
errors in the input quantities; we have taken ad hoc 10\% one-sigma
errors in these models in our orbit systematic error calculations.  At
the consensus distance estimate to 12~Boo these model angular
diameters correspond to model component linear radii of 2.51 $\pm$
0.25 and 1.83 $\pm$ 0.18 R$_{\sun}$ for the primary and secondary
components respectively.  These linear radii values are roughly a
factor of two smaller than the putative Roche lobe radii for these two
stars (\cite{Iben91} Eq.~1), making significant mass transfer unlikely
at this stage of system evolution.  We have further ignored any
corrections due to stellar limb darkening; at these putative component
sizes our data would be highly insensitive to limb-darkening effects
(for examples of observational consequences of stellar limb darkening
see \cite{Quirrenbach96,Hajian98}, and references therein).

\begin{table}
\begin{center}
\begin{small}
\begin{tabular}{ccc}
\hline
Physical	 & Primary           & Secondary \\
Parameter        & Component         & Component \\
\hline \hline
a (10$^{-2}$ AU) & 6.205 $\pm$ 0.032 (0.031/0.008)   & 6.322 $\pm$ 0.046 (0.045/0.008)  \\
Mass (M$_{\sun}$)& 1.435 $\pm$ 0.023 (0.023/0.004)   & 1.408 $\pm$ 0.020 (0.019/0.004)  \\
\cline{2-3}
Sp Type (Barry 1970) & \multicolumn{2}{c}{F9 IVw} \\
System Distance (pc) & \multicolumn{2}{c}{36.93 $\pm$ 0.56 (0.43/0.37)} \\
$\pi_{orb}$ (mas)    & \multicolumn{2}{c}{27.08 $\pm$ 0.41 (0.31/0.27)} \\
\cline{2-3}
Model Diameter (mas) & {\em 0.63} ($\pm$ {\em 0.06}) & {\em 0.46} ($\pm$ {\em 0.05})   \\
M$_{K-{\rm CIT}}$ (mag)      & 1.200 $\pm$ 0.038 (0.031/0.027)   & 1.818 $\pm$ 0.039 (0.032/0.029)  \\
M$_{H-{\rm CIT}}$ (mag)      & 1.269 $\pm$ 0.048 (0.037/0.034)   & 1.835 $\pm$ 0.063 (0.048/0.044)  \\
M$_V$ (mag)      & 2.524 $\pm$ 0.052                 & 3.024 $\pm$ 0.077 \\
$V$-$K$ (mag)    & 1.324 $\pm$ 0.044                 & 1.206 $\pm$ 0.072 \\
\hline
\end{tabular}
\end{small}
\caption{Physical Parameters for 12~Boo.  Summarized here are the
physical parameters for the 12~Boo system as derived primarily from
the Full-Fit solution orbital parameters in Table \ref{tab:orbit}.
Quantities listed in italics (i.e. the component diameters, see text
discussion) are constrained to the listed values in our model fits.
As for all our PTI-derived orbital parameters we have quoted both
total error and separate contributions from statistical and systematic
sources (given as $\sigma_{stat}$/$\sigma_{sys}$), with the exceptions
of quantities involving the spectroscopic $\Delta V$ determination
where the error is assumed to be statistical, and is large compared
with the interferometric estimates.  Infrared absolute magnitudes are
quoted in the CIT system (\cite{Elias82,Elias83}).
\label{tab:physics}}
\end{center}
\end{table}

\section{Component Intensity Ratio}
\label{sec:ratio}
As given in Table \ref{tab:physics}, despite the fact that the mass
ratio of 12~Boo is near 1 ($q \sim$ 0.98), our best fit model of the
system requires that the near-infrared luminosities of the two
components are significantly different.  This brightness difference
appears to be verified in the visible in high-resolution spectroscopy
(see \S \ref{sec:tidal}).

\begin{figure}
\epsscale{0.7}
\plotone{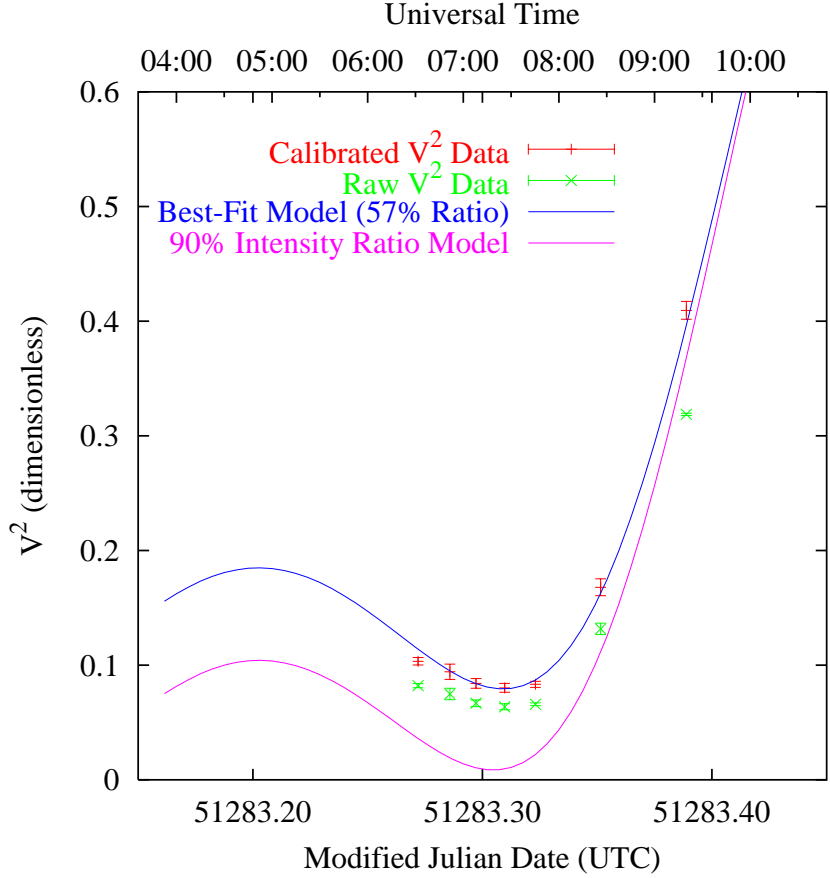}
\caption{12~Boo Visibility Scan Through Minimum.  In order to test the
brightness asymmetry hypothesis, a tentative 12~Boo orbit model was
used to predict a $V^2$ minimum on 15 April 1999.  Both raw and
calibrated $K$-band $V^2$ measurements on 12~Boo are shown, along with
the predictions derived from our ``Full-Fit'' orbit model with a 57\%
component intensity ratio (Table~\ref{tab:orbit}), and the same orbit
but an assumed 90\% component intensity ratio.  The fact that the raw
and calibrated $V^2$ measurements are significantly above the 90\%
intensity ratio model is unequivocal evidence that the $K$-band
component brightness ratio is significantly less that unity.
\label{fig:v2min}}
\end{figure}

The most sensitive test of this relative component brightness
difference is the $V^2$ value at interferometric visibility minima;
following Eq.~\ref{eq:V2_double}, for unresolved components the
minimum $V^2$ on a binary system is approximated by:
\begin{equation}
V^{2}_{min} \approx \left( \frac{1 - r}{1 + r} \right)^2
\label{eq:v2Min}
\end{equation}
In the limit that $r \sim 1$, $V^2 \sim 0$, and a fringe tracking
interferometer like PTI cannot track the source (\cite{Colavita99a}).
Having seen indications of the intensity asymmetry in our initial
orbit solution, on 15 April 1999 we performed a deliberate experiment
to measure the $V^2$ on 12~Boo through a predicted visibility minimum.
The results are given in Figure \ref{fig:v2min}.  Both raw and
calibrated $K$-band $V^2$ measurements on 12~Boo are shown, along with
the predictions derived from our ``Full-Fit'' orbit model with a 57
$\pm$ 1 \% component intensity ratio (Table~\ref{tab:orbit}), and the
same orbit geometry but an assumed 90\% component intensity ratio.
The fact that the raw and calibrated $V^2$ measurements are
significantly above the 90\% intensity ratio model is unequivocal
evidence that the $K$-band component brightness ratio is significantly
less that unity.  Moreover, our best-fit orbit model in general, and
our $K$-band intensity ratio estimate (0.57, Table~\ref{tab:physics})
in particular, does an excellent job of predicting the $V^2$
variations (Eq.~\ref{eq:V2_doubleWB}) and the minimum $V^2$
(Eq.~\ref{eq:v2Min}) respectively.

\begin{figure}
\epsscale{0.7}
\plotone{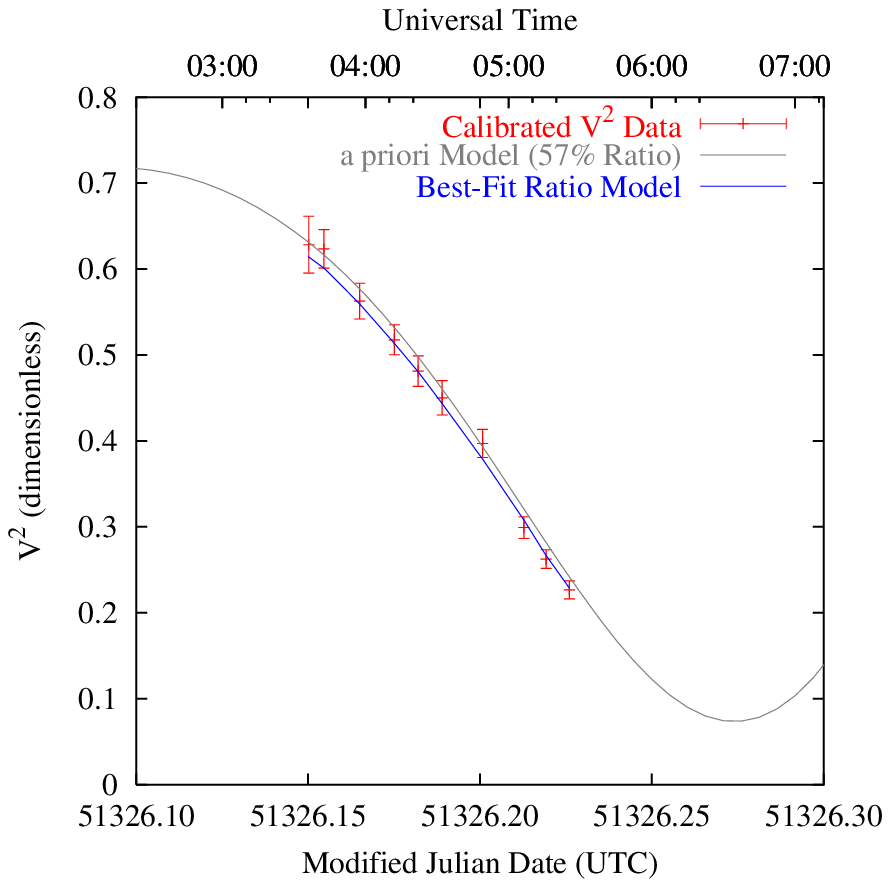}
\caption{12~Boo $H$-band Visibilities.  In order to estimate the
differential $H$-magnitude of the two 12~Boo components we observed
12~Boo on one of our first $H$-band science commissioning runs at PTI
on 28 May 1999.  Our $K$-band derived 12~Boo orbit model, including
the $K$-band component intensity ratio, was used to predict the
expected $V^2$ variations at $H$-band.  A fit to the $H$-band data for
component intensity ratio results in a slightly smaller component
magnitude difference than we derive in our $K$-band fit, but the
difference is not formally significant (see text and Table
\ref{tab:physics}).
\label{fig:v2hband}}
\end{figure}

In 1999 PTI added the capability to make $H$-band ($\lambda \sim$ 1.6
$\mu$m) visibility measurements.  As mentioned in
\S~\ref{sec:observations}, 12 Boo was observed with PTI in $H$-band on
one night (28 May 1999) from which we estimate an $H$-band magnitude
difference of the two components.  Figure \ref{fig:v2hband} shows the
calibrated $H$-band visibilities obtained on 12~Boo, and a priori
(using the $K$-band intensity ratio) and fit predictions for 12~Boo.
This small amount of $H$-band data indicates that the magnitude
difference of the 12~Boo components in $H$ is 0.566 $\pm$ 0.045 (Table
\ref{tab:orbit}); the difference between this and our $K$-band value
of 0.618 $\pm$ 0.021 is not formally significant.

\begin{figure}
\epsscale{0.7}
\plotone{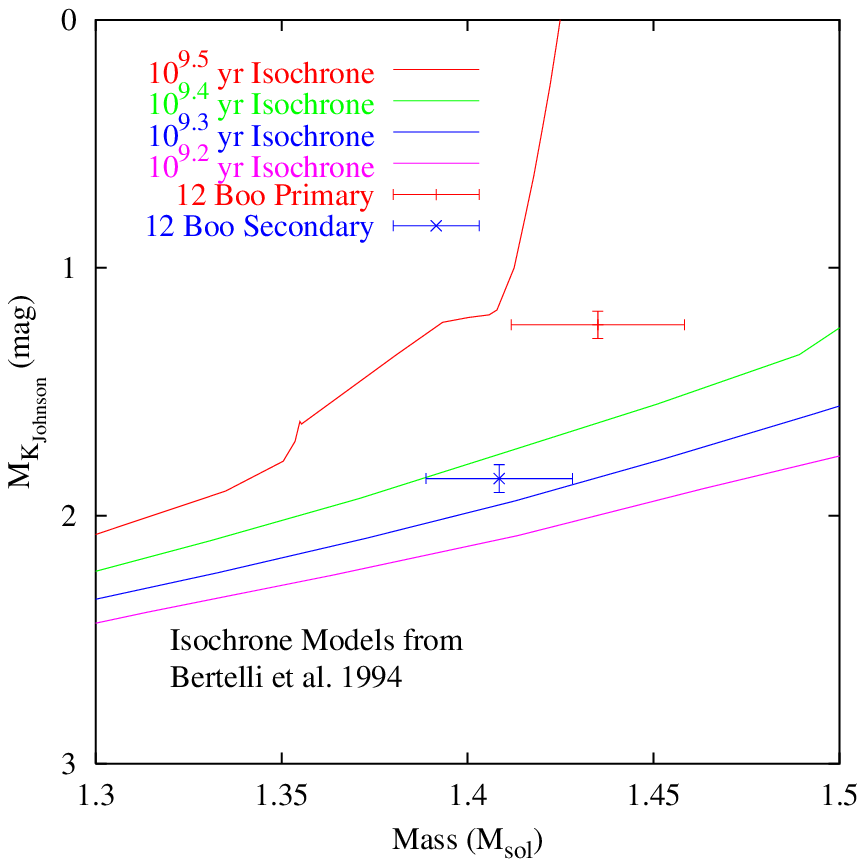}\\
\plotone{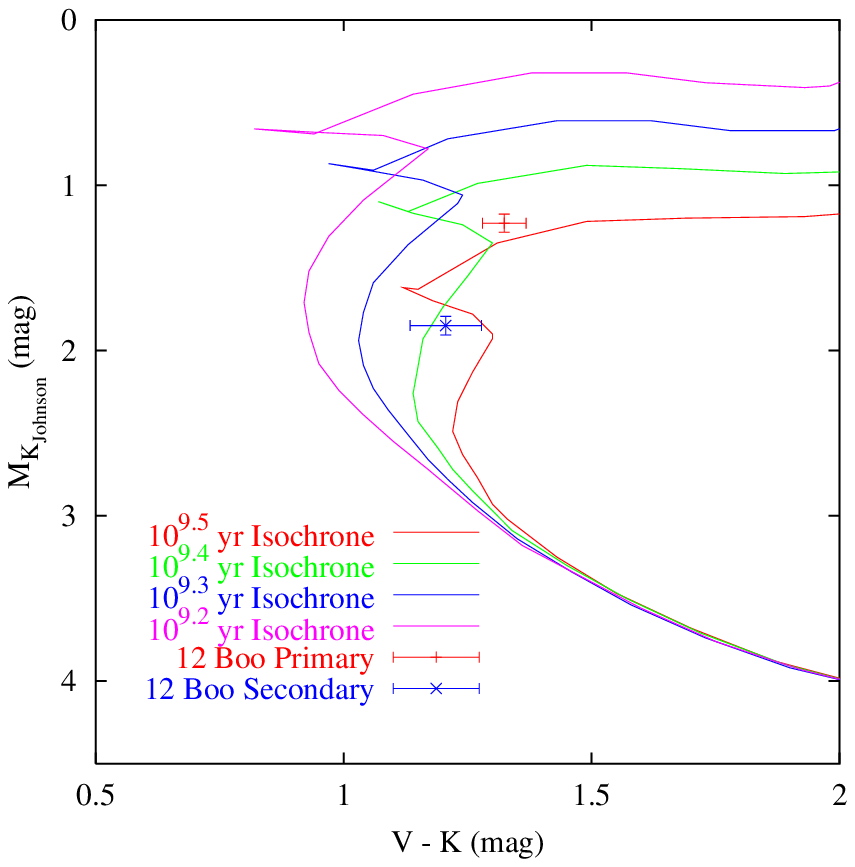}
\caption{12~Boo Components In Mass-Magnitude and Color-Magnitude
Spaces.  Here we depict the 12~Boo components in observable parameter
mass-magnitude and color-magnitude spaces, along with theoretical
isochrone tracks from B94.
\label{fig:isochrones}}
\end{figure}

\paragraph{Comparisons With Stellar Models}
Given our estimates of component masses, absolute magnitudes, and
color indices derived from our measurements and orbital solution
(Table~\ref{tab:physics}), we can examine the 12~Boo components in the
context of stellar models.  Figure~\ref{fig:isochrones} shows the
placement of the 12~Boo components in observable parameter
mass-magnitude and color-magnitude spaces, along with theoretical
isochrone evolutionary tracks from Bertelli et al.~(1994, hereafter
B94).  Here we have used B94 isochrone tracks for $Z$ = 0.02 (B94
Table 5) based on the assumption of solar abundances for the two
12~Boo components.  The [Fe/H] values given in
(\cite{Duncan81,Balachandran90,Lebre99}) indicate $Z_{\rm 12~Boo}$ =
0.0177 $\pm$ 0.0017 (B94 Eq.~10), hence the applicability of the solar
abundance models.  The isochrones suggest that the primary component
of 12~Boo is in the midst of particularly rapid evolution;
unfortunately the coverage of the isochrone models is relatively
coarse in the region of interest for this system.  B94 uses Johnson
infrared passbands in their calculations; for comparison with B94
models we have transformed our component $M_K$ values
(Table~\ref{tab:physics}) from the CIT to the Johnson system using the
relation from Bessell \& Brett (1988).

By jointly fitting our component $M_V$, $M_K$, and $V$ - $K$ estimates
(Table \ref{tab:physics}) to the B94 model predictions of these
quantities at our estimated component masses as a function of age we
find a best-fit 12~Boo system age estimate of 2.550 $\pm$ 0.023 Gyrs.
The quoted error in this age estimate is purely statistical and does
not include possible systematic error effects in the B94 isochrones.
The fit is of relatively poor quality, with a chi-squared of 10.3 for
two degrees of freedom.  If instead we use only the observed component
magnitude differences in $V$ and $K$ (Table \ref{tab:physics})
compared to the B94 models we find a best-fit system age of 3.220
$\pm$ 0.022 Gyrs (again the error is purely statistical).  That these
two different age estimates do not agree within their statistical
errors indicates a significant discrepancy between the B94 models and
our observed parameters for the components in the 12~Boo system.  This
discrepancy makes it difficult to decide which of the two different
estimates is more accurate.  A simple average of these two different
age estimates would suggest a 12~Boo system age of 2.89 $\pm$ 0.36
Gyrs; the error in this age estimate is apparently dominated by
systematic errors in the application of the B94 stellar models to the
12~Boo components.

\section{Tidal Interaction and Component Rotation in 12~Boo}
\label{sec:tidal}
In short-period binary systems, the components gravitationally
interact so as to circularize the orbit and synchronize the component
rotations to the orbit period (\cite{Zahn77,Hut81}).  In a survey of
nearby ``solar-like stars'', Duquennoy \& Mayor (1991) give evidence
that systems with periods shorter than 11 days are ``all circularized
due to tidal effects occurring during their evolution on the main
sequence.''  The circularization and synchronization phenomena
necessarily require an energy dissipation mechanism, which is
generally thought to be associated with convection in the outer
envelopes of evolved stars (\cite{Verbunt95}).

12~Boo is interesting from a tidal interaction perspective because the
system orbit is modestly eccentric ($e$ $\sim$ 0.19;
Table~\ref{tab:orbit}), yet the components of the system have in fact
evolved off the main sequence.  With our masses
(Table~\ref{tab:physics}) and eclipsing binary-derived models
(\cite{Popper80,Andersen91}) we infer that during their main-sequence
lives the 12~Boo components were of approximate spectral type F3 V.
This typing would suggest the reason for the remnant eccentricity in
the 12~Boo system is the putative lack of a convective dissipation
mechanism in early-F main sequence stars.  However, as the components
evolve their atmospheres become more convective, and tidal
circularization and synchronization should begin.

While the orbit remains eccentric, the characteristic timescale for
rotational synchronization in close binaries with convective envelopes
is approximately given by (\cite{Zahn77,Giuricin84}):
\begin{displaymath}
t_{\rm synch} \sim \left(\frac{R}{a}\right)^{-6} \; q^{-2} \; {\rm yr}
\end{displaymath}
with $R$ as the stellar radius, $a$ as the orbital semi-major axis,
and $q$ as the binary mass fraction.  Our models for the 12~Boo
components and orbit imply synchronization timescales of
1.6$\times$10$^6$ and 1.1$\times$10$^7$ yr for the primary and
secondary components respectively.  Because these are significantly
shorter than the likely system age of $\sim$ 3 Gyr
(\S~\ref{sec:ratio}) it is interesting to look for signs of
synchronization in 12~Boo.

Several recent measurements of the rotation \vsini~of 12~Boo exist
(\cite{Balachandran90,DeMedeiros97,Lebre99}, DU99), offering the
possibility to test whether the two components are synchronously
rotating.  Additionally we have taken a high resolution ELODIE
spectrum of the system to assess the rotation and the relative
brightness of the components (see below).  We summarize these recent
rotation measurements and our own in Table~\ref{tab:rotation}.  The
consensus is that both 12~Boo components are rotating considerably
faster than the mean rotational velocity of 5.4 km s$^{-1}$ for
subgiant stars with ($B$-$V$) $\sim$ 0.55 (\cite{DeMedeiros96}).

\begin{table}
\begin{center}
\begin{small}
\begin{tabular}{ccc}
\hline
		 	& Primary           & Secondary \\
	         	& \vsini            & \vsini    \\
			& (km s$^{-1}$)     & (km s$^{-1}$) \\
\hline
\cite{Balachandran90}   & 10 $\pm$ 3	    & 		\\
\cite{DeMedeiros97}	& 12.7 $\pm$ 1      &		\\
\cite{Lebre99}		& 12.7 $\pm$ 1      &		\\
DU99			& 12.5 ($\pm$ 1)    & 9.5 ($\pm$ 1)  \\
ELODIE/This Work        & 13.1 $\pm$ 0.3    & 10.4 $\pm$ 0.3 \\
\hline
Composite 	        & 13.00 $\pm$ 0.28  & 10.33 $\pm$ 0.29 \\
\hline
Model Synch Rotation		& 12.5 ($\pm$ 1.2)  & 9.1 ($\pm$ 0.9) \\
Model Pseudo-Synch Rotation	& 15.2 ($\pm$ 1.5)  & 11.1 ($\pm$ 1.1) \\
\hline
\end{tabular}
\end{small}
\caption{Recent \vsini~Measurements for 12~Boo Components.  Summarized
here are the most recent \vsini~measurements for the 12~Boo system
components.  For references where a single \vsini~measurement is
listed we have assumed this pertains to the primary component.  DU99
does not list errors for their component \vsini~estimates; we have
arbitrarily taken 1 km s$^{-1}$ so as to be consistent with the
characteristic accuracies of earlier CORAVEL determinations (see
discussions in \cite{DeMedeiros96,DeMedeiros97,Lebre99}).  We list
weighted average \vsini~estimates based on the listed measurements.
For comparison we further give model estimates of \vsini~for
synchronous and pseudo-synchronous rotation of the two components
assuming the physical sizes discussed in \S~\ref{sec:physics}; errors
are based on ad-hoc estimates of systematic errors in the diameter
models.
\label{tab:rotation}}
\end{center}
\end{table}

\begin{figure}
\epsscale{0.8}
\plotone{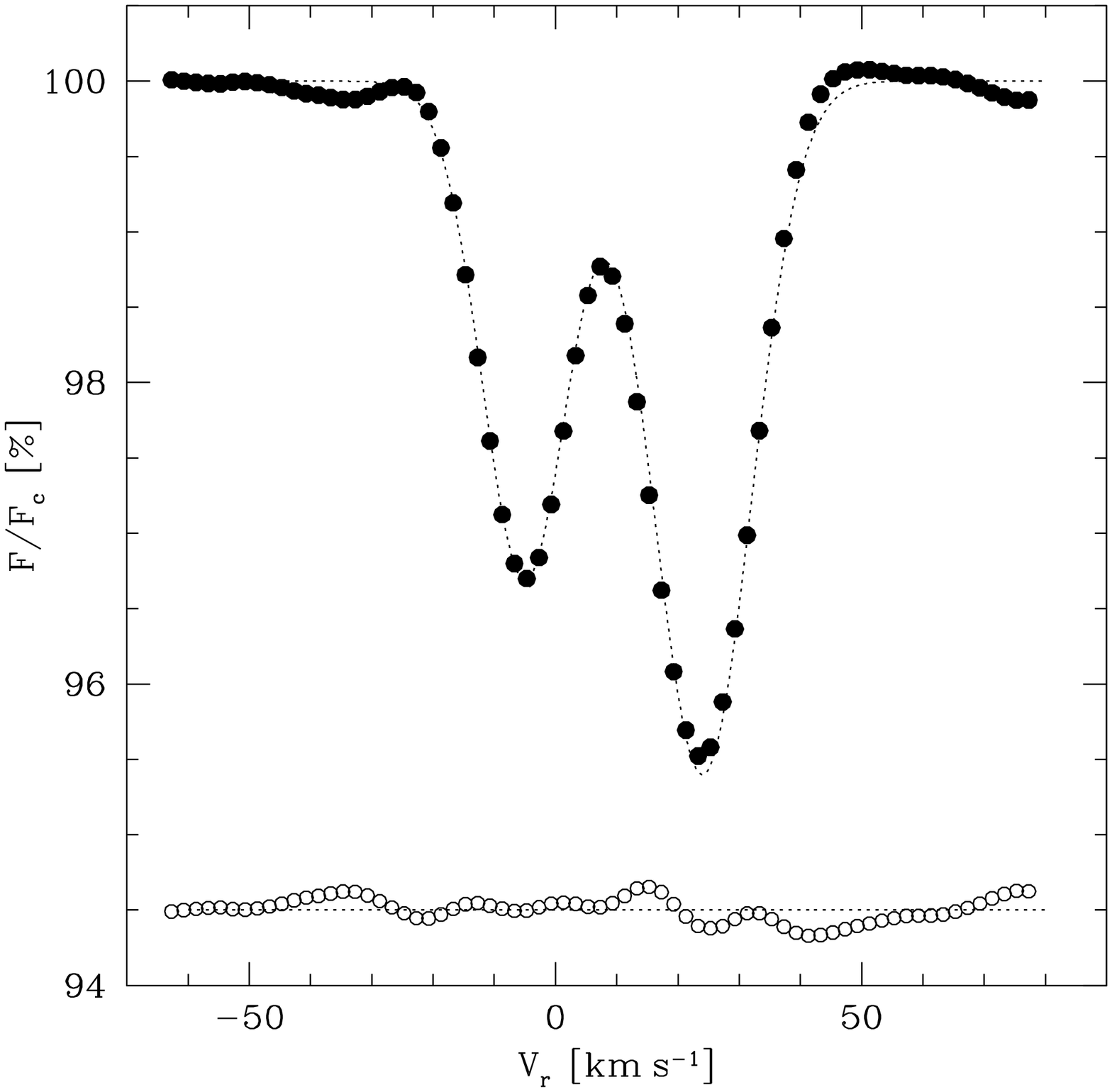}
\caption{Cross-correlation function of the 12~Boo ELODIE spectrum
(filled dots). We see the double line feature characteristic of a
double-lined spectroscopic binary system. The superimposed dotted line
is the double Gaussian fit.  Empty dots at the bottom display the
residuals ($\sigma=0.07$\% rms).
\label{fig:crossCorrFigure}}
\end{figure}

\paragraph{ELODIE Measurements}
In June 1999 we took a spectrum of the 12~Boo system with the ELODIE
high resolution echelle spectrograph at the Observatoire de Haute
Provence, in France (\cite{Baranne96}).  A cross-correlation function
(CCF) has been computed using the on-line reduction package available
with this instrument (Figure \ref{fig:crossCorrFigure}).

The CCF corresponds to a mean of the spectral lines selected by the
template (\cite{Queloz94}).  The width of the CCF is a measurement of
the mean broadening of spectral lines.  It can be calibrated for each
spectral type to provide a \vsini~measurement of the star
(\cite{Queloz98}).  The equivalent width of the CCF ($W$) is sensitive
to the temperature and the metallicity of the star.  If the stellar
spectral lines have no strong asymmetries or the star is not a fast
rotator the shape of the CCF can be very well approximated by a
Gaussian function ($G(v)$) in absorption, ($1-G(v)$).

In a double-lined spectroscopic binary such as 12~Boo, the equivalent
width ratio ($W_1/W_2$) can be used as a flux ratio indicator if the
intrinsic equivalent width ratio ($(W_1/W_2)^{\rm intrisic}$) can be
estimated from two similar single spectra: $W_i=W_i^{\rm intrisic}\,
F_i (\sum_j F_j)^{-1}$, then $W_1/W_2=(W_1/W_2)^{\rm intrisic}\,
(F_1/F_2)$.  The mass of each component in the 12~Boo system is very
similar (Table \ref{tab:physics}).  The CCF equivalent width ratio
computed using respectively only the blue and the red part of the
ELODIE spectrum show no significant differences, suggesting very
similar $T_{\rm eff}$.  Since the system is coeval a metallicity
difference between the two objects seems unlikely.  Therefore we make
the assumption that the intrinsic equivalent width of each of the
spectral line systems is similar, and use our computed CCF as an
indicator of relative brightness.  With this assumption a magnitude
difference $\Delta m_v = 0.5 \pm 0.1$ is found, in reasonable
agreement with the component magnitude difference seen in the infrared
interferometric data (\S~\ref{sec:ratio}).  The derived intrinsic
equivalent width agrees with a $(B-V)\approx0.55$ star with solar
metallicity.  Note that the gravity difference expected between the
two stars is too small to have a significant impact on the spectral
line equivalent widths.

To compute the \vsini~of each of the stars from the CCF, the Queloz et
al.~(1998) calibration has been used. This leads to \vsini$=13.1 \pm
0.3$ km s$^{-1}$ for the bright (primary) component and \vsini$=10.4
\pm 0.3$ km s$^{-1}$ for the dimmer (secondary) component
(Table~\ref{tab:rotation}).  Note that the smaller gravity of the
brighter component compared to the other one desaturates some lines
from the bright object.  This would cause an underestimate of \vsini,
however the effect is small in our analysis, and well within the 0.3
km s$^{-1}$ error in our measurements.

Assuming the 12~Boo component model diameters calculated above
(\S~\ref{sec:physics}), and coplanarity of orbital motion and
component spin (required for tidal equilibrium), we can calculate the
expected \vsini~speeds for synchronous or pseudo-synchronous rotation
(where the rotation equilibrates with the orbital motion at periapsis
when the tidal forces are at a maximum (\cite{Hut81})) with the
orbital motion; these are listed in Table~\ref{tab:rotation}.  The
quoted errors in these rotation calculations are dominated by the ad
hoc uncertainties in our component model diameters.  Our ELODIE
\vsini~measurements suggest that both components are rotating slightly
faster than synchronously, but neither component's \vsini~seems
consistent with the possibility of pseudo-synchronous rotation.
However, we would add that the DU99 \vsini~values lead one to conclude
that both components are in essentially synchronous rotation.  Because
DU99 do not list errors in their \vsini~values it is difficult to
assess the relation between our measurements and theirs.

Finally, in an attempt to independently confirm component rotation
periods through differential photometry we consulted the Automated
Astronomy Group at Tennessee State University.  Both the Phoenix-10
automatic photoelectric telescope (APT) in Phoenix, AZ, and the
Vanderbilt/Tennessee State 16-inch APT at Fairborn Observatory near
Washington Camp, AZ, have routinely observed 12~Boo as a photometric
reference star, and thus have a large body of differential photometric
data on the system.  They report $V$-band night-to-night scatter at
the 0.004 mag level (consistent with scatter on other photometric
references), and no discernible modulation associated with component
rotation (\cite{Henry99}).

\section{Summary and Discussion}
By virtue of our interferometric resolution and the precision of the
AL76 and CORAVEL radial velocity data we are able to determine
accurate physical parameters for the 12~Boo constituents, and an
accurate system distance.  Our 12~Boo distance estimate is in
excellent agreement with the Hipparcos trigonometric determination.
Our finding of unexpectedly large relative $K$, $H$, and $V$-magnitude
differences in the two nearly-equal mass 12~Boo components suggests
that the system is in a unique evolutionary state, with both
components making the transition off the main sequence.

The agreement between our findings on the physical parameters of the
two 12~Boo components and the stellar models of B94 is not
particularly good.  The fundamental reason for the discrepancy is the
unexpectedly large magnitude difference we see in the interferometric
and spectroscopic data; clearly it leads one to question the veracity
of the data in this regard.  As described in \S~\ref{sec:ratio}, for
the interferometric data we performed dedicated observations looking
to verify the component brightness asymmetry.  We have further looked
in detail at the individual spectral channel data for both the
standard PTI pipeline processing, and for alternate detector
calibrations (\cite{Colavita99b}).  In all cases the conclusion is the
same: the interferometric visibility measurements remain significantly
above zero through the visibility minima.  Unless there is some
unknown and heretofore unseen positive bias in our visibility
processing at low visibilities (such effects should have been evident
in \cite{Boden99b} in particular), the conclusion of a component
brightness asymmetry seems sound.  Having seen the brightness
difference in the near-infrared, we verified this conclusion with
visible spectroscopy as discussed in \S~\ref{sec:tidal}.

Accepting the component brightness difference as genuine, and further
assuming that the component metallicities are similar and the stars
coeval leads us to the conclusion that the dominant source of error in
the age estimate for the 12~Boo system arises in B94's treatment of
stars at this stage of evolution.  The system metallicity is
reasonably well-determined (\cite{Duncan81,Balachandran90,Lebre99}),
and our spectroscopy shows no indications of dissimilar metallicities
between the components.  While one might expect possible metallicity
differences in an interacting binary system, computations indicate
that neither component of the 12~Boo system has filled its Roche lobe
(\S~\ref{sec:physics}), making it unlikely that significant mass
transfer has occurred to date in the system.  The calculated
separation of the two components is approximately 0.12 AU, a strong
suggestion that these stars formed from the same molecular cloud.
Finally, there is no indication in the IRAS mid-IR photometric
measurements for the 12 Boo system of an infrared excess, which might
indicate the presence of dust (as in a common envelope or mass-loss
from one of the components).

We believe the two most likely reasons for the discrepancy between our
component parameters (Table \ref{tab:physics}) and the B94 model
tracks stem from the B94 treatment of convection.  Iben (1991)
indicates that the time during which a 1.5 M$_{\odot}$ star
initially moves to the right of the main sequence on an H-R diagram
(before the hook back to the left) is the time when hydrogen is
burning in a convective core.  Near the approximate derived age of the
12~Boo system ($\log\; T$ = 9.4 yr, 2.51 Gyr), B94 note that there is
a dip in the relation between the absolute magnitude and age for the
turn-off from the main sequence (at the point where the star
transitions from radiative to convective core H-burning).  Further,
while B94 do treat overshooting from convective cores in their models,
only two empirical values (based on measurements of galactic clusters)
are used for $\Lambda_c$, 0.25 for
1.0~M$_{\odot}$~$\leq$~M~$\leq$~1.5~M$_{\odot}$ and 0.5 for
M~$\geq$~1.6~M$_{\odot}$.  The fact that the 12~Boo system age is
likely close to $\log\; T$ = 9.4 yrs, and that the component masses
are close to the 1.5~M$_{\odot}$ B94 transition point in $\Lambda_c$
is suggestive that the B94 models may be too coarse in this region of
parameter space to describe the particular stage of component
evolution seen in the 12~Boo system.  Conversely, we are optimistic
that the physical parameters derived here for the 12~Boo components
are of sufficient quality to service as significant constraints to
evolutionary modeling.

One potential additional contributor to the data-model discrepancy is
the possibility of biases in our mass estimates.  While we see good
statistical agreement between the DU99 and our ``Full-Fit'' solution
(e.g.~1.2-sigma in the component semi-amplitudes), our solution calls
for a smaller component mass difference than the DU99 solution would
indicate.  For example, were we simply to adopt the DU99
semi-amplitudes and our ``$V^2$-only'' astrometric orbit we would
obtain component mass estimates of 1.464 $\pm$ 0.031 and 1.403 $\pm$
0.028 M$_{\sun}$ for the primary and secondary components
respectively.  These alternative mass values are statistically
consistent with our ``Full-Fit'' values (Table \ref{tab:physics}), and
would alleviate some, but not all, of the discrepancy with the B94
models (see Figure \ref{fig:isochrones}a in particular).  The reason
for the potential mass-bias in our results is clear: our inclusion of
the AL76 RV data, which led AL76 to in fact conclude that the 12~Boo
secondary was more massive than the primary.

Although the short-period orbit of 12~Boo has not yet circularized,
tidal interactions should be at work in the system as both components
become more convective with their evolution off the main sequence.
The body of \vsini~measurements suggests that both components of
12~Boo are rotating at or slightly higher than synchronous rotational
speed, and significantly faster than the mean rotational speed for
similar subgiant stars (\cite{DeMedeiros96}).  Taken at face value the
\vsini~measurements suggest that the primary is closer to synchronous
rotation than the secondary; this seems consistent with the rough
factor of 7 difference in the synchronization time scale expected from
the model size difference between the two components.

\acknowledgements The work described in this paper was performed at
the Infrared Processing and Analysis Center, California Institute of
Technology, and the Jet Propulsion Laboratory under contract with the
National Aeronautics and Space Administration.  Interferometer data
were obtained at Palomar Observatory using the NASA Palomar Testbed
Interferometer, supported by NASA contracts to the Jet Propulsion
Laboratory.  Science operations with PTI are conducted through the
efforts of the PTI Collaboration
(http://huey.jpl.nasa.gov/palomar/ptimembers.html), and we acknowledge
the invaluable contributions of our PTI colleagues.  Photometric data
were obtained with the 200'' Hale telescope at Palomar Observatory,
operated by the California Institute of Technology.  We thank
K.Y.~Matthews (CIT) for assistance with near-infrared photometry on
12~Boo.  Spectroscopic data were obtained at the Observatorie de
Haute-Provance with ELODIE at the 193 cm telescope, and with CORAVEL
at the 1 m Swiss telescope; we acknowledge the gracious cooperation of
the Geneva Extrasolar Planet Search Programme.  A.F.B.~in particular
thanks C.A.~Hummel (USNO) for suggestions concerning integrated
fitting of interferometric visibilities and radial velocities, and
G.~Henry (Tennessee State University) for his timely analysis of APT
photometry on 12~Boo.  We further thank G.~Bertelli for his gracious
consultation on the application of the B94 models.  And we thank the
anonymous referee for his many helpful suggestions on improving the
quality of this manuscript.

This research has made use of the Simbad database, operated at CDS,
Strasbourg, France.


\begin{thebibliography}{99}

\bibitem[Abt \& Levy 1976]{Abt76}
Abt, H.~ and Levy, S.~1976 (AL76), \apjs~30, 273.

\bibitem[Andersen 1991]{Andersen91}
Andersen, J.~1991, \aapr~3, 91.

\bibitem[Armstrong et al.~1992]{Armstrong92b}
Armstrong, J.T.~et al.~1992, \aj~104, 2217.

\bibitem[Balachandran 1990]{Balachandran90}
Balachandran, S.~1990, \apj~354, 310.

\bibitem[Baranne et al.~1996]{Baranne96}
Baranne A., Queloz D., Mayor M., Adriansyk G., Knispel G., et al.~1996, 
\aaps~119, 1.

\bibitem[Barry 1970]{Barry70}
Barry, D.~1970, \apjs~19, 281.

\bibitem[Bertelli et al.~1994]{Bertelli94}
Bertelli, G., Bressan, A., Chiosi, C., Fagotto, F., and Nasi, E.~1994 (B94), \aaps~106, 275.

\bibitem[Bessell \& Brett~1988]{Bessell1988}
Bessell, M.~and Brett, J.~1988, \pasp~100, 1134.

\bibitem[Blackwell et al.~1990]{Blackwell90}
Blackwell, D., Petford, A., Arribas, S., Haddock, D., and Selby, M.~1990, \aaps~232, 396.

\bibitem[Blackwell \& Lynas-Gray 1994]{Blackwell94}
Blackwell, D., and Lynas-Gray, A.~1994, \aap~282, 899.

\bibitem[Boden et al.~1998]{Boden98}
Boden, A.F.~et al.~1998, \procspie~3350, 872.

\bibitem[Boden et al.~1999a]{Boden99a}
Boden, A.F.~et al.~1999a, \apj~515, 356 (astro-ph/9811029).

\bibitem[Boden et al.~1999b]{Boden99b}
Boden, A.F.~et al.~1999b, \apj~in press (astro-ph/9905207).

\bibitem[Campbell \& Wright]{Campbell00}
Campbell, W.W.~and Wright, W.H.~1900, \apj~12, 254.

\bibitem[Colavita et al.~1999]{Colavita99a}
Colavita, M.M.~et al.~1999, \apj~510, 505 (astro-ph/9810262).

\bibitem[Colavita 1999]{Colavita99b}
Colavita, M.~1999, \pasp~111, 111 (astro-ph/9810462).


\bibitem[De Medeiros et al.~1996]{DeMedeiros96}
De Medeiros, J.R., Da Rocha, C., and Mayor, M.~1996, \aap~314, 499.

\bibitem[De Medeiros et al.~1997]{DeMedeiros97}
De Medeiros, J.R., Do Nascimento, J., and Mayor, M.~1997, \aap~317, 701.

\bibitem[De Medeiros \& Udry 1999]{DeMedeiros99}
De Medeiros, J.R., and Udry, S.~1999 (DU99), \aap~346, 532.

\bibitem[Duncan 1981]{Duncan81}
Duncan, D.K.~1981, \apj~248, 651.

\bibitem[Duquennoy \& Mayor 1991]{Duquennoy91}
Duquennoy, A., and Mayor, M.~1991, \aap~248, 485.

\bibitem[Elias et al.~1982]{Elias82}
Elias, J., Frogel, J., Matthews, K., and Neugebauer, G.~1982, \aj~87, 1029.

\bibitem[Elias et al.~1983]{Elias83}
Elias, J., Frogel, J., Hyland, A., and Jones, T.~1983, \aj~88, 1027.

\bibitem[ESA~1997]{HIP97}
ESA 1997, The Hipparcos and Tycho Catalogues, ESA SP-1200.

\bibitem[Giuricin et al.~1984]{Giuricin84}
Giuricin, G., Mardirossian, F., and Mezzetti, M.~1984, \aap~141, 227.

\bibitem[Hajian et al.~1998]{Hajian98}
Hajian, A.~et al.~1998, \apj~496, 484.

\bibitem[Henry 1999]{Henry99}
Henry, G.~1999, Private comm.

\bibitem[Hummel et al.~1993]{Hummel93}
Hummel, C.A.~et al.~1993, \aj~106, 2486.

\bibitem[Hummel et al.~1995]{Hummel95}
Hummel, C.~et al.~1995, \aj~110, 376.

\bibitem[Hummel et al.~1998]{Hummel98}
Hummel, C.~et al.~1998, \aj~116, 2536.

\bibitem[Hut 1981]{Hut81}
Hut, P.~1981, \aap~99, 126.

\bibitem[Iben 1991]{Iben91}
Iben, I.~1991, \apjs~76, 55.

\bibitem[Lebre et al.~1999]{Lebre99}
Lebre, A., De Laverny, P., De Medeiros, J., Charbonnel, C., and Da Silva, L.~1999,
\aap~345, 936.

\bibitem[Merrill 1922]{Merrill22}
Merrill, P.~1922, \apj~56, 40.

\bibitem[McAlister, Hartkopf, \& Mason 1992]{McAlister92}
McAlister, H., Hartkopf, W., and Mason, B.~1992, \aj~104, 1961.

\bibitem[Mozurkewich et al.~1991]{Mozurkewich91}
Mozurkewich, D.~et al.~1991, \aj~101, 2207.

\bibitem[Perryman et al.~1997]{Perryman97}
Perryman, M.A.C.~et al.~1997, \aap~323, L49.

\bibitem[Popper 1980]{Popper80}
Popper, D.~1980, \araa~18, 115.

\bibitem[Press et al.~1992]{Press92}
Press, W.H., Teukolsky, S.A., Vetterling, W.T., and Flannery, B.P.~1992,
Numerical Recipes in C: The Art of Scientific Computing, Second Edition,
Cambridge University Press.

\bibitem[Queloz 1994]{Queloz94}
Queloz D.~1994, in IAU 167 "New Developments in Array Technology 
and Applications", Ed.~A.G.~Davis Philip, 221.

\bibitem[Queloz et al.~1998]{Queloz98}
Queloz D., Allain S., Mermillod J.-C., Bouvier J., Mayor M.~1998,
\aap~335, 183.

\bibitem[Quirrenbach et al.~1996]{Quirrenbach96}
Quirrenbach, A.~et al.~1996, \aap~312, 160.

\bibitem[Selby et al.~1988]{Selby88}
Selby, M.~et al.~1988, \aaps~74, 127.

\bibitem[Tokovinin 1997]{Tokovinin97}
Tokovinin, A.~1997, \aaps~124, 75.

\bibitem[Verbunt \& Phinney 1995]{Verbunt95}
Verbunt, F.~and Phinney, E.~1995, \aap~296, 709.

\bibitem[Worley \& Douglass 1997]{WDS97}
Worley, C.~and Douglass, G.~1997, \aaps~125, 523.

\bibitem[Zahn 1977]{Zahn77}
Zahn, J.-P.~1977, \aap~57, 383.

\end{thebibliography}
\end{document}